\documentclass[showpacs,preprintnumbers,amsmath,amssymb,twocolumn]{revtex4}
\usepackage[final]{graphicx}
\usepackage[english]{babel}

\begin{document}
\title{
 Collisional energy losses of heavy quarks in relativistic nuclear
    collisions at energies available at the BNL Relativistic Heavy Ion
    Collider}
\author{Yu.A. Tarasov}
\email{tarasov@dni.polyn.kiae.su}
\affiliation{
 Russian National Research
Center ''Kurchatov Institute'', 123182, Moscow,  Russia }

\date{\today}

\begin{abstract}
We investigate the  energy losses of $b$ and $c$ quarks at hard collisions with gluons
 and light quarks in quark-gluon plasma produced in central  $Au + Au$  collisions
 at RHIC energy($\sqrt{s_{NN}} =  200$ GeV). As  the coupling constant at hard collision in
the perturbative QCD  is limited by Landau pole at momenta $Q^{2} = \Lambda^{2}$,
we use the analytic models of QCD, where the Landau pole is absent.  We calculate
the energy losses of $b$ and $c$ quarks in analytic model of QCD  at $Q^{2} \ge \Lambda^{2}$.
At calculations we use the effective quasiparticle model. We use at calculations of
total nuclear modification factor $R_{AA}^{b+c}$  the fragmentation of $b$ and $c$ quarks into
heavy hadrons with accounting theirs energy losses. We show that this factor gives the
considerably suppression of heavy $B$ and $D$ mesons in the region of middle meanings of $p_{\perp}$.
The  experimental data~\cite{2} show, that nonphotonic electrons have the analogous suppression
in the same region of $p_{\perp}$.

\end{abstract}


\maketitle

\section{Introduction    \label{Sec.1}}

Heavy quarks, unlikely light quarks do not come to an equilibrium with the
surrounding matter and may therefore play an important role in the search
for properties of this matter. Produced in hard collisions, their initial
momentum can be inferred from pp collisions. The deviation of the measured
heavy mesons $p_{\perp}$ distribution in $AA$ collisions from that measured in pp collisions is usually quantified by the  nuclear modification factor
\begin{equation}
\label{eq.1}
R_{AA} = \frac{dN_{Au+Au}}{T_{AA}\,d\sigma_{p+p}},
\end{equation}
where $dN_{Au+Au}$ is the differential electrons yield from heavy meson decays in $Au+Au$ collisions and
$d\sigma_{p+p}$ is corresponding cross section in $p+p$ collisions in any given $p_{\perp}$ bin.

In the RHIC experiments the heavy mesons have not yet been directly measured. Both the PHENIX~\cite{1} and STAR~\cite{2} collaboration have
observed only single nonphotonic electrons, which have been created in the
semileptonic decay of heavy mesons. Thus experimentally it is not easy
separate between charm and bottom hadrons. The QCD calculations in the
fixed order + next to leading logarithm (FONLL) predict~\cite{3} the
theoretical uncertainty for the charm and bottom quarks $p_{\perp}$
distribution in $pp$-collisions at $\sqrt{s_{NN}}=200$ GeV. Apparently, that
above $p_{\perp} = 4$ GeV/c the electrons from bottom mesons dominate the
spectrum. However, uncertainty here is considerable. The nonphotonic
electron yield at RHIC measurement exhibits unexpectedly large suppression
in central $Au + Au$ collisions at high $p_{\perp}$. The observed values
$R_{AA} \sim 0.2$ are smaller than originally expected. The  theoretical
approaches based on perturbative QCD (pQCD) calculation give much larger
values and it has been doubt, whether pQCD is the right description of this
interaction. One should note that for light quarks the radiative energy
loss is the most important. The heavy quarks have a smaller amount of
radiative energy loss due to their large mass~\cite{4},~\cite{5}.

The asymptotic freedom in QCD enable one to apply the standard perturbative
formulas at large momenta transferred. However a number of phenomena are
beyond such calculation. Moreover, the perturbative theory contain unphysical singularities in the expression for the running coupling, for example the ghost poles and also the unphysical cutes. Therefore by number
of authors were considered the analytic models of invariant charge in QCD.
The condition of analyticity follows from the general principles of local
Quantum Field Theory. The analytic models contain no unphysical singularities and running coupling may be determined in the spacelike
(Euclidean) region and in the timelike region by analytic continuation
from spacelike region. In this work we investigate the collisional energy
losses of $b$ and $c$ quark in quark-gluon plasma (QGP) produced in central
$Au+Au$ collisions at RHIC energy. We use here the formulas for analytic
invariant charge in QCD in the spacelike region. We show that the nuclear
modification factor $R_{AA}^{b+c}$ observed in the experiments is determined by hard collisions with QCD analytic charge $\alpha_{an}(Q^2)$ which have the Landau pole at $Q^{2}\cong\Lambda^2$ in the perturbative QCD, but have no of this pole
in analytic  QCD. The vicinity $Q^{2}\cong \Lambda^{2}$ one should consider as boundary of hard collisions in QCD.

In Sec.~\ref{Sec.2} we give the brief description of analytic models results, which we use in this paper, and corresponding reference.

In Sec.~\ref{Sec.3} we give the brief description of effective quasiparticle model and physical characteristic of expanding plasma. We calculate here
energy losses $\Delta E/E$ of $b$ and $c$ quarks at collisions with
gluons and light quarks in plasma, using the analytic coupling in QCD. We
calculate also the summary losses of $b$ and $c$ quarks for various energies
while they transverse the expanding plasma.

In Sec.~\ref{Sec.4} we calculate the common nuclear modification factor
$R_{AA}^{b+c}$ for $b$ and $c$ quarks, which fragmentized into $B$ and $D$ mesons.
We use the fragmentation function $b$ and $c$ quarks with accounting of theirs
energy losses. We take into account the relative $B$ and $D$ contribution into
electrons from theirs semileptonic decays. We compare this modification
factor with experimental data.

In Sec.~\ref{Sec.5} - Conclusion.

\section{ Analytic models for the QCD running coupling \label{Sec.2}}

The effective coupling constant $\alpha_{s}(Q^2)$ is given in the lowest
approximation (one loop) by the famous asymptotic formulas
\begin{equation}
\label{eq.2}
\alpha_s(Q^2) = \frac{4\pi}{(11- \frac{2N_f}{3})\ln\frac{Q^2}{\Lambda^2}},
\end{equation}
where $\Lambda$ is dimensional constant of QCD.
This coupling constant have divergence at $Q^2 = \Lambda^{2}$. That is
related to an infrared Landau pole in the running coupling, that is
nonperturbative issue. In the two-loop case we have besides pole also the
unphysical cut owing to the factor of type $\ln(\ln\frac{Q^2}{\Lambda^2})$.
In consequence of these difficulties the number of authors use the analytic models of invariant charge in QCD. For example, some of them; ~\cite{6},~\cite{7},~\cite{8},~\cite{9}. The K\"allen - Lehman spectral representation
satisfied to condition of analyticity and it applied to the "analytization" of the renormalization group (RG) equations. We describe here briefly the
analytic model proposed in the works~\cite{8},~\cite{9}. Here the solution is derived as the solution of the RG equation with the $\beta$ function
"improved" by the analytization procedure. This correspond to demand of
analyticity of right hand side of RG equations as a whole before their
solution by use of the K\"allen - Lehmann integral representation. At the
$l$-loop level the analytic running coupling $\alpha_{an}^{(l)}(q^2)$ is
defined as the solution of equation:
\begin{equation}
\label{eq.3}
\frac{d\ln\alpha_{an}^{(l)}(q^2)}{d\ln q^2} =
- \{\sum\limits_{j=0}^{l-1}B_{j}[\alpha_{s}^{(l)}(q^2)]^{j+1}\}_{an}.
\end{equation}

The value $\alpha_{an}^{(1)}(q^2)$ at the one-loop level has here the form:
\begin{equation}
\label{eq.4}
\alpha_{an}^{( 1)}(q^2) = \frac{4\pi(z-1)}{\beta_0z\ln z},
\end{equation}
where $z = \frac{q^2}{\Lambda^2}$, $\beta_{0} = 11 - \frac{2n_{f}}{3}$
This invariant charge have no of the Landau pole in the vicinity of the
point $z=1$, namely $\alpha_{an}^{(1)}$($z=1+\epsilon)\simeq \frac{4\pi}{\beta_{0}}(1-\frac{\epsilon}{2})$ and it describe also the asymptotic for large $z$. The infrared enhancement at small $z$ is in agreement with the  Schwinger - Dyson equations~\cite{10}.
The analytic invariant charge (AIC) at the higher loop levels was also
investigated in the works~\cite{8},~\cite{9} on the basic of the
K\"allen - Lehmann type integral representation. It was shown that AIC
possesses the higher loop stability. i.e. the curves corresponding to
the two and three-loop levels are almost indistinguishable.
In the present work we use analytic invariant charge in the form~(\ref{eq.4}) for investigation of energy losses $b$ and $c$
quarks at collisions with gluons and light quarks in the quark-gluon
plasma produced in central $Au+Au$ collisions at RHIC energy.

\section{Collisional energy losses of $b$ and $c$ quarks in quark-gluon
plasma \label{Sec.3}}

By analogy with the works~\cite{11},~\cite{12} we use the effective
quasiparticle model for calculation of physical characteristics of
expanding plasma. We assume the hot glue production at the early stage.
This is caused by the relatively large $gg$ cross section in comparison
with $qg$ and $qq$ ones~\cite{13}. In the work~\cite{12} were calculated
the initial temperature $T_{0}$ and time $\tau_{0}$ at $\sqrt{s_{NN}} =200$ GeV and also $\tau(T)$ for evolution of the plasma phase for $T\to T_{c}$ from above. For these estimations we used the formulas of the quasiparticle model for gluons and quarks densities $n_{g}$, $n_{q}$, $n_{s}$ at
temperature $T$ at expanding of plasma~\cite{14}. These densities depend from thermal
masses $m_{g}(T)$ and $m_{q}(T)$ of gluons and quarks. These thermal masses are connected
with the thermodynamic functions of medium and they are determined by effective
coupling $G^{2}(T)$. They describe well the lattice data at $ T_{c}\le T \le 4T_{c}$.
We use here the parametrization of effective two-loop coupling $G^{2}(T,T_{c})$ from
Ref.~\cite{14}. We investigate in this work the central $Au+Au$ collisions for nuclei
modeled as a uniform cylinder with sharp edge and effective radius $R_{eff}\simeq
6$ fm, which correspond to $\tau\simeq 6$ fm and $T\simeq$ 183 MeV~\cite{15}. This
value of $T$ is close to $T_{c}=177$ MeV~\cite{16}.
The calculation of the lepton spectrum includes initial $b$ and $c$ quarks distribution
from perturbative QCD, heavy  quarks $Q$ energy losses, their fragmentation into
heavy mesons $H_{Q}$ and  electron decay  spectrum of heavy mesons. This cross section
is schematically written as~\cite{3},~\cite{5}:
\begin{eqnarray}
\label{eq.5}
\frac {Ed^{3}\sigma(e)}{dp^3} &=& \frac{E_{i}d^{3}\sigma(Q)}{dp_{i}^{3}} \bigotimes P(E_{i} \to E_{f})\bigotimes \nonumber \\ &&\bigotimes  D(Q \to H_{Q}) \bigotimes f(H_{Q} \to e),
\end{eqnarray}
where the symbol $\bigotimes$ denotes a generic convolution. The term $f(H_{Q} \to e)$
includes the branching ratio to electrons.
The initial $b$ and $c$~$p_{\perp}$ distributions in $pp$ collisions with account of
theoretical uncertainty  were computed at FONLL level in Ref.~\cite{3}. In this work
we use these differential cross section $\frac{d\sigma}{dydp_{\perp}}$ (look at Fig.~\ref{Fig.1}).

The invariant cross section for collisions of heavy quark $Q$ with momentum ${\textbf p}$
and massless partons with momentum $\textbf k$  in plasma can be written in the form:
\begin{eqnarray}
\label{eq.6}
d\sigma &=& d{\textbf p'}d{\textbf k'}\frac{4\alpha_{s}^{2}}{4E_{p'}k'} \frac{|M|^{2}}{4J} \delta(\textbf p+\textbf k -\textbf p' -\textbf k') \times \nonumber \\ &&\times \delta(E_{p} + |k| - E_{p'} - |k'|) \nonumber \\  &&\equiv d{\textbf p'}d{\textbf k'} F({\textbf p},{\textbf k},{\textbf p'},{\textbf k'}, m_{Q}),
\end{eqnarray}
where ${\textbf p'}$ and ${\textbf k'}$ are the momenta of outgoing heavy quarks $Q$ and light partons in plasma, $|M|^{2}$ is the matrix element squared. The value $J = \sqrt{(PK)^2}$, where $PK=E_p |k| - {\textbf p}{\textbf k}=k (E_p-p\cos{\Psi})$ and $\Psi$ the angle between vectors {\textbf p} and {\textbf k}. Thus $J=\frac{S-M_Q^2}{2}$, where $S=M_Q^2+2|k|(E_p-p\cos{\Psi})$.
Using the formula~(\ref{eq.6}) for cross section $d\sigma$,
one should determine the relative energy loss $\frac{\Delta E}{Ed\tau}$  per unit length
(for $Qg$ collisions) from integral:
\begin{equation}
\label{eq.7}
\int \frac{d{\textbf k}\,n_{g}({\textbf k})}{(2\pi)^3} \frac{\omega}{p}\, d\sigma,
\end{equation}
where $\omega = E_{p} - E_{p'}$ (in $t$ channel) and $n_{g}(\textbf k)$ is the gluon
density in the quasiparticle model. We calculate in these equations at first the
$k'$-integral, using three-dimensional $\delta$ function, and this gives
$k'=|{\textbf k}+{\textbf p}-{\textbf p'}|$. For the second $\delta$ function it is
convenient to choose for massless partons the $z$ axis along the direction ${\textbf p}$
of the heavy quarks, and ${\textbf k}$ contained in the $yz$ plane~\cite{12},~\cite{17}.
We use the second $\delta$ function for calculation of the integral $I$ over azimuthal angle $\phi$ of the form:
\begin{equation}
\label{eq.8}
I = \int_0^{2\pi}\delta\left(E_{p}-E_{p'}+k-\sqrt{A+B\cos\phi}\right).
\end{equation}

The result of calculation is:
\begin{equation}
\label{eq.9}
I = \frac{2}{\sqrt{B^2 -A^2}}.
\end{equation}

In values $A$ and $B$ we change the variables from $p'$ and $\cos\theta$ to variables $t$ and $\omega$ with corresponding
Jacobian, and $t = 2M_{Q}^2 - 2(E_{p}E_{p'} - pp'\cos\theta$). The formula of type~(\ref{eq.9}) was used in Ref.~\cite{17} for investigation of collisional energy loss of muons with mass $M$ in the hot QED plasma. The value
$B^2 - A^2$ has here  the form: $B^2 - A^2 = - a^2 \omega^2 + b\omega + c$. We use in this paper the coefficients
$a$, $b$, $c$ from Ref.~\cite{17} for the masses $M_{Q}$ of heavy quarks. The value $B^2 - A^2$ is positive in interval
$\omega_{\min} \le \omega \le \omega_{\max}$, where $\omega_{\min}^{\max} = \frac{b\pm\sqrt{D}}{2a^2}$. The
 discriminant $D = 4a^{2}c + b^2$ is~\cite{17}:
 \begin{equation}
 \label{eq.10}
D = -t[st+(s - M_{Q}^2)^2]\left(\frac{4k\sin\Psi}{p}\right)^2.
\end{equation}

The condition $D\ge 0 $ gives $0\le |t| \le\frac{(s-M_{Q}^{2})^2}{s}$. However
we use here the analytic model of QCD with lower boundary of |t| in the vicinity of Landau
pole $t\simeq\Lambda^{2}$ where the QCD running coupling for hard collisions contain no
unphysical singularity. We can  evaluate now the relative energy loss $\frac{\Delta E}{E}$
for $b$ and $c$ quarks using the equations~(\ref{eq.6}) and~(\ref{eq.7}) and the $\omega$ integral:
\begin{equation}
\label{eq.11}
\int\limits_{\omega_{\min}}^{\omega_{\max}} \frac{\omega d\omega}{\sqrt{B^2 - A^2}} =
\frac{\pi|t|p[E_p(s - M_{Q}^2) - k(s + M_{Q}^2)]}{(s - M_{Q}^2)^3}.
\end{equation}
We have thus the formula (for $Qg$ collisions):
\begin{eqnarray}
\label{eq.12}
\frac{\Delta E}{E d\tau} = &&\int\frac{d {\textbf k}\, n_{g}({\textbf k})}{(2\pi)^3} \int\limits_{|t|_{\min}}^{(s - M^2)^2/s} d |t| \, \alpha_{ac}^{2}(t) \pi |t| \times \nonumber \\ &\times& \frac{(s + M^2)\left[\frac{E(s - M^2)}{s + M^2} - k\right]
}{(s - M^2)^{3} E}
 \frac{|M_{Qg}|^2}{4J},
\end{eqnarray}
where $4J = 2(s - M^2)$, and $M$ is $M_b$ or $M_c$ - the mass of $b$ or $c$ quark. The square of the matrix element in
$t$ channel have the form~\cite{18}: $\frac{|M_{Qg}|^2}{4J} = \frac{\overline{s} - |t|}{t^2}$ where
$\overline{s} =  s - M^2$. In quasiparticle model the thermal distribution of gluons is
$n_{g}(k) = \frac{1}{\exp{(\sqrt{y^2 + \frac{m_{g}^{2}(T)}{T^2})}-1}}$, where $k = yT$, and the thermal mass $m_{g}(T)$ is
expressed across the effective two - loop coupling $G^{2}(T,T_c)$ from Ref.~\cite{14}. From equation~(\ref{eq.12}) we
have the condition  $(E - \frac{M^2}{E - p\cos\Psi}) \ge k$. It is convenient to introduce the variable
$\xi_1 = E/p - \cos\Psi$, i.e. the condition is $E - \frac{M^2}{\xi_{1}p} \ge k$. We have also $\overline{s} = 2yT \xi_{1}p$. From this condition and also from the condition $\frac{\overline{s}^2}{s} \ge |t|_{\min}$ we have
$k \ge \frac{\sqrt{|t|_{\min}} M}{2\xi_{1}p} \left(1+\frac{\sqrt{t_{\min}}}{2M}\right)$ and $\xi_{1} \ge \frac{M^2}{Ep}\left(1+\frac{\sqrt{|t|_{\min}}}{2M}\right)$. The upper limit for variable $|t|$ in formula~(\ref{eq.12}) is
\begin{equation}
\label{eq.13}
\frac{(\overline{s})^2}{s}=\frac{(2y\xi_{1})^{2} T^{2}\frac{p^2}{M^2}}{1+2y\xi_{1}\frac{Tp}{M^2}}.
\end{equation}

We use  for calculations the mass of heavy quarks $M_{b}=4.75$ GeV, $M_{c}=1.3$ GeV. For example, at $p = 8$ GeV we
have $E_{b} = 9.304$ GeV and $E_{c} = 8.1$ GeV. We have also $\xi_{1\max} = (E/p +1)$.
We write down now the formula~(\ref{eq.12}) for heavy quarks with the mass $M$ with accounting of analytic invariant
charge in the form~(\ref{eq.4}):
\begin{eqnarray}
\label{eq.14}
\frac{\Delta E}{E d\tau} &=&   \frac{16 T}{4\pi}\frac{T^2}{pE} \left(\frac{4\pi}{\beta_0}\right)^2 \times \nonumber \\ &\times& \int\limits_{\xi_{1\min}}^{E/p +1} \frac{d\xi_{1}}{2\xi_{1}} \int\limits_{y_{\min}}^{\frac{E}{T} \left(1-\frac{M^2}{pE\xi_{1}}\right)} dy \, y \frac{\left[\frac{E}{T}\left(1-\frac{M^2}{pE\xi_{1}}\right) -y\right]}{e^{\sqrt{y^2 - \frac{m_g^2(T)}{T^2}}} -1} \nonumber \\ && \int\limits_{\frac{t_{\min}}{\Lambda^2}}^{\frac{(2y\xi_{1})^2\frac{T^{2}p^2}{\Lambda^{2}M^2}}{1 + 2y\xi_{1}\frac{Tp}{M^2}}}
\frac{dx}{z} \left(\frac{z-1}{z\ln z}\right)^{2},
\end{eqnarray}
where $\xi_{1\min} = \frac{M^2}{Ep}(1+\frac{\sqrt{t_{\min}}}{2M})$, $y_{\min} \simeq \frac{M\sqrt{t_{\min}}}{2\xi_{1}pT}$. We use here the analytic approach in QCD for investigate of energy losses of heavy quarks at hard collisions with light partons in plasma at temperature $T_0\ge T\ge T_c$. We investigate the energy losses of $b$ and $c$ quarks in the region of hard collisions  $|t| \ge \Lambda^{2}$, where the perturbative QCD is no applicable in the vicinity of Landau pole $|t| \simeq \Lambda^{2}$.

 In the work~\cite{12} were found the physical
parameters $T$ and $\tau$ for expanding plasma at $\sqrt{s}= 200 A$ GeV with initial meanings $T_0 = 410.6$ MeV and
$\tau_{0} = 0.405 m_{\pi}^{-1}$. We use here the number of this parameters for calculations of $\frac{\Delta E}{E}$ for
$b$ and $c$ quarks at different energies by formula~(\ref{eq.14}). These calculations  were fulfilled by means of numerically.  We show in the Table~\ref{Tab.1} the energy losses of $b$ and $c$ quarks for $bg$ and $cg$ collisions for series of $T$ and $\tau$ for expanding plasma (for example at $p_{b,c} = 8$ GeV). We have here at $|t|_{min} = \Lambda^2$ for $b$ quark
($M_b = 4.75$ GeV) $\xi_{1\min}= 0.31$, $y_{\min} =\frac{0.0064 E_b}{\xi_{1}T}$, and for $c$ quark ($M_c = 1.3$ GeV)
$\xi_{1\min} = 0.028$, and $y_{\min} = \frac{0.002 E_c}{\xi_{1}T}$, and for Landau pole $z_{\min} = 1$.
The invariant analytic charge~(\ref{eq.4}) in formula~(\ref{eq.14}) describe also the asymptotic for large $z$, and we use
the ordinary QCD constant $\Lambda = 200$ MeV.
In the third  column of the Table~\ref{Tab.1} we give $\frac{\Delta E}{E}$ for $bg$ collisions, in the fourth column we give the same energy losses for $cg$, and then we give the summary energy losses $(\frac{\Delta E}{E})_{tot}$.

\begin{table}
\caption{The relative energy losses for $b$ and $c$ quark at $bg$ and $cg$ collisions}
\label{Tab.1}
\begin{tabular}{cccc}
 \hline    \hline
 $T$(MeV)  &  $\tau$($m_{\pi}^{-1})$ &  $(\frac{\Delta E}{E})_{b}$  &  $(\frac{\Delta E}{E})_{c}$
 \\ \hline
410.6      &    0.405     &  0.143     &   0.49
\\
390        &     0.474    &   0.128     &   0.433
\\
370         &     0.543    &   0.113     &   0.379
\\
350         &      0.613    &   0.0993    &   0.329
\\
330         &      0.691    &    0.086    &    0.283
\\
310         &      0.825    &   0.0738    &    0.239
\\
290         &      0.98     &   0.0618    &    0.199
\\
270         &      1.07     &   0.0508    &    0.162
\\
250         &      1.19     &   0.0404    &    0.127
\\
240         &      1.26     &   0.0355    &     0.115
\\
233         &      1.31     &    0.0324   &      0.1
\\
220         &      1.62     &    0.026    &     0.0806
\\
210         &      1.92     &    0.0211   &     0.0656
\\
200         &      2.36     &    0.0163   &     0.0505
 \\
190         &      3.08     &    0.0112    &     0.0347
\\
183         &      4.13     &    0.0073     &     0.0226
\\                \hline        \hline
\end{tabular}
\end{table}

The calculation gives for summary energy losses: for $b$ quark $(\frac{\Delta E}{E})_{tot}^{b} = 0.112$, and for $c$ quark
$(\frac{\Delta E}{E})_{tot}^{c} = 0.357$.

One should determine also the energy losses of heavy quarks at collisions with light quarks in the plasma. We assume
that in the initial stage at $T = T_{0}$ all the quarks are nonequilibrated. With decrease of temperature (at $T \le T_{0})$
the number of equilibrated quarks increase. In the work~\cite{12} we estimated the part $\beta$ of equilibrated quarks
at expansion of plasma. At $ T \le $ 233 MeV the all quarks become equilibrated. We estimate here the heavy quarks
energy losses also at collisions with portion of nonequilibrated quarks. We assume  that nascent quarks have the similar
momentum distribution as gluons~\cite{13}. The matrix element squared $|M|^2$ have the factor 4/9 for $bq$ and $cq$ collisions, however there is the some exceeding of the number of quarks over gluons one at middle temperature. We
show in the Table~\ref{Tab.2} the results of energy losses calculation $\frac{\Delta E}{E}$ for $bq$ and $cq$ collisions
by analogy with the Table~\ref{Tab.1} also for $p_{b,c} = 8$ GeV.

\begin{table}
\caption{The relative energy losses for $b$ and $c$ quarks at $bq$ and $cq$ collisions}
\label{Tab.2}
\begin{tabular}{cccc}
\hline    \hline
$T$ (MeV)  &   $\tau$ $(m_{\pi}^{-1})$   &   $(\frac{\Delta}{E})_{b}$   &   $(\frac{\Delta}{E})_{c}$
\\ \hline
410.6    &   0.405  &  0.0295    &   0.1013
\\
390      &    0.474  &  0.036    &    0.1222
\\
370      &    0.543   &  0.0402   &    0.1351
\\
350      &    0.613   &  0.0454   &     0.1511
\\
330      &    0.691   &   0.0502   &     0.165
\\
310      &     0.825   &  0.055    &     0.1803
\\
290      &     0.98    &   0.0612   &    0.198
\\
270      &     1.07    &    0.051   &     0.1627
\\
250      &     1.19    &    0.042   &     0.133
\\
240      &     1.26    &    0.0381  &     0.12
\\
233      &     1.31    &    0.0348   &     0.1085
\\
220      &     1.62     &   0.029    &      0.0903
\\
210      &     1.92     &    0.025   &      0.0776
\\
200      &     2.36     &    0.0205   &     0.0635
\\
190      &     3.08     &     0.0156  &     0.0484
\\
183      &     4.13     &     0.0113   &    0.035
\\           \hline        \hline
\end{tabular}
\end{table}

The calculation gives for summary energy losses: for $b$ quark $(\frac{\Delta E}{E})_{tot}^{b} = 0.097$ and for
c quark $(\frac{\Delta E}{E})_{tot}^{c} = 0.309$.

In this work we calculated   the summary  energy losses $(\frac{\Delta}{E})_{tot}$ of $b$ and $c$ quarks for various energies. However in Sec~\ref{Sec.4}  we use the fragmentation function for $b$ and $c$ quark into heavy $B$ and $D$ mesons with accounting of energy losses $b$ and $c$ quarks. Therefore one should evaluate the energy losses in relative values
$\frac{\Delta p}{p}$ for various $p$ with accounting of heavy quark masses $M$. It is convenient for this to use the
values $\frac{\Delta E}{p}$, where $\Delta E = \sqrt{p^2 + M^2} - \sqrt{p'^{2} + M^2}$. We have with accounting
of the next term $\frac{M^4}{p^4}$:
\begin{equation}
\label{eq.15}
\frac{\Delta E}{p} = \frac{\Delta p}{p} + \frac{M^2}{2p^2}\left(1 - \frac{1}{1 - \frac{\Delta p}{p}}\right) - \frac{M^4}{8 p^4} \left[1 - \frac{1}{\left(1 - \frac{\Delta p}{p}\right)^3}\right].
\end{equation}

Using   the calculation of summary values $(\frac{\Delta E}{p})_{tot}$ for various values $p$ and formula~(\ref{eq.20}), we show in Fig.~\ref{Fig.2} the results of numeral calculations of $\frac{\Delta p}{p}(p)$ for $bg$, $cg$ and $bq$, $cq$ collisions ($M_b = 4.75$ GeV, $M_c = 1.3$ GeV). These results we use in Sec.~\ref{Sec.4} for calculation of nuclear
modification factor $R_{AA}$.

\section{Calculation of nuclear modification factor $R_{AA}$ for $b$ and $c$ quarks \label{Sec.4}}

The heavy quarks $b$ and $c$ fragmentize into heavy $B$ and $D$ mesons (look at scheme~(\ref{eq.5})). In the PHENIX~\cite{1} and
STAR~\cite{2} experiments are observed only single nonphotonic electrons in the semileptonic decays of heavy mesons.
These fragmentation functions were considered in the number works, for example ~\cite{19},~\cite{20}, ~\cite{21} and
others. The fragmentation function contain usually the different parameters for $b$ and $c$ quarks. If the energy  of
heavy quark decrease before fragmentation, its transverse momentum is shifted by energy loss on the value $\Delta p(p)$. This effect consist in replace of fragmentation  function $D(z)$ by effective one $\frac{z'}{z} D(z')$~\cite{22},
where:
\begin{equation}
\label{eq.16}
z' = \frac{z}{1 - \frac{\Delta p}{p}},
\end{equation}
where  $z = \frac{p_{H}}{p_{Q}}$. We use here the Peterson fragmentation function~\cite{19}, which has been used
most extensively in applications:
\begin{equation}
\label{eq.17}
D_{Q}^{H}(z) = \frac{N_{H} z (1-z)^{2}}{[(1-z)^{2} + \epsilon_{Q} z]^{2}}.
\end{equation}

The parameter $N_{H}$ is normalization, and $\epsilon_{Q}$ is $\sim \frac{m_{q}^2}{m_{Q}^2}$, the ratio of the effective
light and heavy quark masses. Usually is used the traditional values $\epsilon_{b} \simeq 0.006$ and $\epsilon_{c} \simeq 0.06$~\cite{23}. However the value  $\epsilon_{b} = 0.006$ is appropriate only when a leading-log (LL)
calculation of spectra is used. When NLL (the next to leading level) calculation are used, smaller values of
$\epsilon$ are needed to fit the data (for bottom quarks cross section). In the work~\cite{24} is used the more
appropriate value $\epsilon_{b} = 0.002$ (with FONLL calculation) to fit data. In the present work we use the
fragmentation function~(\ref{eq.17}) with this parameter $\epsilon_{b} = 0.002$. It should be noted also that parameter
$\epsilon_{c} = 0.06$ is extracted from LO fits to charm production data. However this parameter extracted from NLO
fits is $\epsilon_{c}\simeq 0.02$~\cite{25}. This parameter $\epsilon_{c} = 0.02$ we use also in the fragmentation function~(\ref{eq.17}). The spectra of heavy mesons $H$ after of fragmentation of $Q$ quark is determined by formula:
\begin{equation}
\label{eq.18}
\frac{d\sigma_{1}}{dp_{\perp}^{H}} = \int\limits_{z_{\min}}^{1} dz \frac{D_{Q}^{H}(z)}{z} \frac{d\sigma}{dp_{\perp}^{Q}},
\end{equation}
 where $p^{Q} = p^{H}/z$. The heavy quarks distributions $\frac{d\sigma}{dp_{\perp}^{Q}dy}$ for $b$ and $c$ quarks
 are shown in Fig.~\ref{Fig.1}. If the heavy quark $Q$ lose $\Delta p (p)$ before fragmentation, the spectra of heavy
 mesons is determined by formula:
 \begin{equation}
 \label{eq.19}
  \frac{d\sigma_{2}}{dp_{\perp}^{H}}  = \int\limits_{z_{\min}}^{z_{\max}} \frac{dz}{z}  \frac{D_{Q}^{H}(z')}{1 - \frac{\Delta p}{p}(z)} \frac{d\sigma}{dp_{\perp}^{Q}}.
\end{equation}

The semi-leptonic decays of heavy mesons dominate the electrons spectrum at RHIC experiment up to $p_{\perp}\simeq 20$ GeV, and we use the value $z_{\min}\simeq \frac{p_{\perp}^{H}}{20}$ for every  meanings of $p_{\perp}^{H}$.
The value $z_{\max}$ is determined from condition $z' = \frac{p_{H}}{p - \Delta p} = \frac{z}{1 - \frac{\Delta p}{p}} = 1$. Using this condition and Fig.~\ref{Fig.2} for $\frac{\Delta p}{p}$, one can build the functions $p_{b}(p_{B})$ and
$p_{c}(p_{D})$ and to find  $z_{\max}$ for various meanings of $p_{B}$ and  $p_{D}$. As we are interested in the total factor $R_{AA}^{b+c}(p_{H})$, one should determine the values $z_{\max}$ for the same momentum $p_{B}$ and $p_{D}$ of heavy mesons.
We  write down for example the several values of $z_{\max}$:

a) for $bg$ and $cg$ collisions:\\
$P_{B} = 6$ GeV    $z_{\max} = 0.857$\\
$p_{D} = 6$ GeV    $z_{\max} =  0.658$\\
$p_{B}  = 6.9$ GeV  $z_{\max}  = 0.86$\\
$p_{D}  = 6.9$ GeV  $z_{\max}  = 0.678$\\
$p_{B}  = 8.5$  GeV   $z_{\max}  = 0.85$\\
$p_{D}  = 8.5$  GeV   $z_{\max}  = 0.71$

b) for  $bq$  and  $cq$  collisions:\\
$p_{B} =  6$ GeV     $z_{\max}$  =  0.88\\
$p_{D} =  6$ GeV     $z_{\max}$  =  0.69\\
$p_{B}  =  6.9$ GeV  $z_{\max}  =  0.864$\\
$p_{D}  =  6.9$ GeV  $z_{\max}  =  0.701$\\
$p_{B}  =  8.5$ GeV  $z_{\max}  =   0.867$\\
$p_{D}  =  8.5$ GeV  $z_{\max}  =   0.74$

The spectra of heavy mesons $H$ after of fragmentation without and with energy losses of heavy quarks is determined
by formulas~(\ref{eq.18}) and~(\ref{eq.19}) with improving parameters $\epsilon_{b} = 0.002$ and $\epsilon_{c} = 0.02$
of Petersons function. The normalization factors $N_{B}$ and $N_{D}$ is fixed  by the probability that $b$ quark fragment
into $B$ meson~\cite{19}: $\int\limits_{z_{\min}}^{1}dz D_{b}^{B}(z) = 1$ and by analogy for $D_{c}^{D}$. Also  the normalization factors $N_{B}'$ and $N_{D}'$ for modified fragmentation functions is fixed by condition~\cite{22}:
\begin{equation}
\label{eq.20}
\int\limits_{z_{\min}}^{z_{\max}^{b}} dz \frac{D_{b}^{B}(z')}{1 - \frac{\Delta p}{p}(z)} = 1
\end{equation}
and similarly for $D_{c}^{D}(z')$. However it should be noted that for example the all $b$ quarks with energy losses
at $bg$  and  $bq$  collisions come from the same spectra $\frac{d\sigma}{dp_{b}}$, therefore one should fix the normalization by condition:
\begin{equation}
\label{eq.21}
  N_{B}' \!\! \left[\int\limits_{z_{\min}}^{z_{\max}^{bg}}\!\!\! dz \frac{D_{b}^{B}(z')}{1 - \left(\frac{\Delta p}{p}\right)_{bg}(z)} + \int\limits_{z_{\min}}^{z_{\max}^{bq}}\!\!\! dz \frac{D_{b}^{B}(z')}{1 - \left(\frac{\Delta p}{p}\right)_{bq}(z)}\right]  =  1.
\end{equation}

We write down here the function $D_{b}^{B}(z')$ without normalization factor $N_{B}'$,
and from here we determine normalization factor $N_{B}'$. From analogous condition for $c$ quarks (with $D_{c}^{D}(z')$ and
corresponding values $z_{\max}^{g,q}$ we determine the factor $N_{D}'$). These conditions correspond to value of the
nuclear modification factor without energy losses - $R_{AA}^{b+c} = 1$.  Thus the total factor $R_{AA}^{b+c}(p_{\perp}^{H})$ for $b$ and $c$ quarks with energy losses it is convenient to write in the form:
\begin{equation}
\label{eq.22}
R_{AA}^{b+c}(p_{\perp}^{H}) =  \frac{N_{B}'(A_{bg} + A_{bq}) + N_{D}'(A_{cg} + A_{cq})}{N_{B}A_{b} + N_{D}A_{c}}.
\end{equation}

We  use here the designation in numerator:
$$ A_{bg} = \int\limits_{z_{\min}}^{z_{\max}^{bg}} \frac{dz}{z} \frac{D_{b}^{B}(z')}{1 - \left(\frac{\Delta p}{p}\right)_{bg}(z)} \frac{d\sigma}{dp_{\perp}^{b}},
$$  and  $A_{bq}$ correspond  to $A_{bg \to bq}$.

We use also: $$ A_{cg} =
\int\limits_{z_{\min}}^{z_{\max}^{cg}} \frac{dz}{z} \frac{D_{c}^{D}(z')}{1 - \left(\frac{\Delta p}{p}\right)_{cg}}
 %\frac{\Delta p}{p})_{cg}(z)}
  \frac{d\sigma}{dp_{\perp}^{c}},
$$  and  $A_{cq}$ correspond to $A_{cg \to cq}$.

The designation in denominator is $$
A_{b} = \int\limits_{z_{\min}}^{1} \frac{dz}{z} D_{b}(z) \frac{d\sigma}{dp_{\perp}^{b}}
$$  and  $$A_{c} = \int\limits_{z_{\min}}^{1} \frac{dz}{z} D_{c}(z)\frac{d\sigma}{dp_{\perp}^{c}}.
$$

In these formulas the fragmentation functions $D_{b}$ and $D_{c}$ we write down also without normalization factors.

However in experiments are observed only non-photonic electrons in the semileptonic decays of heavy mesons.
It was noted~\cite{3}, that apparently the electrons from $b$ mesons  dominate the spectrum for $p_{\perp} \ge 4$ GeV.  But recently the contribution of $B$ meson decays to non-photonic electrons  in $p+p$ collisions at $\sqrt = 200 A$ GeV has been measured using azimuthal correlation between non-photonic electrons and hadrons~\cite{26}.
The extracted $B$ decay contribution is approximately 0.5 at $p_{\perp} \ge 5$ GeV/c. The $\frac{e_{B}}{e_{B} + e_{D}}$
ratio is found to be $0.52 \pm 0.03$. We use this result  for calculation of factor $R_{AA}^{b+c}$ by formula~(\ref{eq.22}),
assuming the absence of the exceeding of electrons spectrum from $B$ mesons.
We use the formulas~(\ref{eq.21}), and~(\ref{eq.22}) for calculation of factors $N_{B}'$ and $N_{D}'$  and also of nuclear
modification factor $R_{AA}^{b+c}(p_{\perp}^{H})$ . We use the Figures~\ref{Fig.1} and ~\ref{Fig.2} for numeral
calculation on the $z$ integrals for every of fixed meanings $p_{\perp}^{B} = p_{\perp}^{D}$. We calculate also the
corresponding factors $N_{B}$ and $N_{D}$ without energy losses. We show here the results of numeral calculation of factor
$R_{AA}^{b+c}(p_{\perp})^{H}$ for example at $p_{\perp}^{B} = p_{\perp}^{D} = 6.9$ GeV/c in accordance with designations in formula~(\ref{eq.22}) for every member in this formula:
$N_{B}' =\frac{1}{14.55+14.147}$,
$A_{bg} + A_{bq} = 0.172 + 0.175$,
 $N_{D}' = \frac{1}{2.646 + 2.742}$,
$A_{cg} + A_{cq} = 0.02 + 0.025$,
and  $N_{B} = \frac{1}{14.312}$,
$A_{b} = 0.277$,
$N_{D} = \frac{1}{3.377}$, $A_{c} = 0.18$.

In result we have the nuclear modification factor for $p_{\perp}^{H} = 6.9$ GeV/c  from formula~(\ref{eq.22}):
$R_{AA}^{b+c}(p_{\perp}^{H} = 6.9$ GeV/c) $= 0.281$

We give in the Table~\ref{Tab.3} the results calculations of factor $R_{AA}^{b+c}$ for some meaning $p_{\perp}^{H}$ at
collisions of $b$ and $c$ quarks with gluons only ($R_{AA}^{Qg}$) and with gluons and light quarks($R_{AA}^{Q(g+q)}$)
\begin{table}
\caption{The nuclear modification factor for collision $b$ and $c$ quark with gluons only and with gluons and light quarks}\label{Tab.3}
\begin{tabular}{ccc}
\hline  \hline
$p_{\perp}^{H}$ (GeV/c) & $R{AA}^{Qg}$ & $R_{AA}^{Q(g+q)}$
\\ \hline
5.5    &   0.22   &   0.235
\\
6      &   0.235   &   0.251
\\
6.9    &   0.265   &   0.281
\\
8      &    0.295   &   0.315
\\
8.5    &    0.315   &   0.335
\\   \hline  \hline
\end{tabular}
\end{table}

We see that accounting of $b$ and $c$ quark collisions also with light quarks gives exceeding of value $R_{AA}^{b+c}$ in
less than 10 per cent. We show in Fig.~\ref{Fig.3} the comparison of nuclear modification factor $R_{AA}^{b+c}(p_{\perp}^{H})$ at 5.5 GeV/c $\le p_{\perp}^{H}\le 8.5$ GeV/c in the present model with experimental data  for central
$Au + Au$ collisions at $\sqrt{s} = 200\, A$  GeV~\cite{2}. We have agreement in afore-cited region of $p_{\perp}^{H}$.
The not great increase of $R_{AA}^{b+c}$ with increase of $p_{\perp}^{H}$ apparently is also in experiment. We do not
calculate this factor $R_{AA}^{b+c}$ for too low meanings of $p_{\perp}^{H}$, where the Petersons  fragmentation
functions in form~(\ref{eq.17}) apparently inapplicable~\cite{19}.

\section{Conclusion  \label{Sec.5}}

We investigate in this work the collision energy losses of $b$ and $c$ quarks at hard collisions with gluons and
light quarks in QGP produced in central  $Au + Au$  collisions at  RHIC energy.  We assume that heavy quarks
not come to an equilibrium  with partons in  plasma. The  hard  collisions  are described  by  formulas  of
perturbative  QCD (pQCD) at large enough momenta transferred. The hard collisions in  QCD  limited  by momenta
$Q^{2}\simeq \Lambda^{2}$, where in the pQCD we have Landau pole in the coupling constant  $\alpha_{s}(Q^2)$ at $Q^2=\Lambda^2$ (the $\Lambda \simeq 200$ MeV is dimensional constant of QCD).  In this work we use the analytic models for the
running coupling in  QCD,  which satisfied  to conditions of analyticity  in  local Quantum  Field  Theory (look at
Sec.~\ref{Sec.2}). We use here the analytic model~\cite{8},~\cite{9} with invariant charge in the form~(\ref{eq.4}),
which have no Landau pole and describe also the asymptotic at large momenta transferred $Q^{2} > \Lambda^{2}$.
In Sec.~\ref{Sec.3} we calculate the energy losses of $b$ and $c$ quarks at hard collisions with light partons at momenta
transferred $Q^{2} \ge \Lambda^{2}$.  We show in Fig.~\ref{Fig.2} the results of calculations of summary losses
$\Delta p/p$ of $b$ and $c$ quarks with momentum  $p_{b,c}$ at hard collisions with gluons and also with light quarks. These results we use in Sec.~\ref{Sec.4} for calculation of the total nuclear modification factor $R_{AA}^{b+c}$. We use  the fragmentation function for $b$ and $c$ quarks into $B$ and $D$ mesons with accounting of summary losses
$\Delta p/p$ and also fragmentation functions without energy losses. We use here the Petersons fragmentation
function~\cite{19}, which has been widely adopted in analysis determining the "hardness" of heavy quarks fragmentation
functions. Its attraction being that it has only one free parameter, $\epsilon_{Q}$, for each heavy quark $Q$. But we
use the Petersons fragmentation functions with improving parameters $\epsilon_{Q}$ for $b$ and $c$ quarks to fit the data, as has been mentioned in Sec.~\ref{Sec.4}.  We take into account to the equal degree the energy losses of $b$ and $c$ quark at
 calculations. One should not neglect by $b$ quark contribution, thought it is less than $c$ quark one.
 We take into account also the relative contribution of $B$ decays $\simeq 50$ percents in non-photonic electrons~\cite{26}.
 We show in Fig~\ref{Fig.3} the comparison of factor $R_{AA}^{b+c}$ for heavy mesons B and D in the present model with experimental
 data~\cite{2}. These data show the analogous suppression of nonphotonic electrons in the region of middle meanings of $p_{\perp}$.
 It is
 interesting to compare the energy losses of heavy quarks and nuclear modification factor with others models.
 We plan also the analogous calculation with others fragmentation functions.

\begin{acknowledgements}
The authors are grateful to Dr.\ A.\ V.\ Lomonosov for a careful reading of the
manuscript. The work was supported in part by the Ministry of the Russian Federation for Science and Higher Education within the project for support
of leading scientific schools (grant NS-7235.2010.2).
\end{acknowledgements}

\newpage
\begin{figure*}
\caption {The differential cross section of charm (the upper curve) and bottom (the lower curve) quarks     in $pp$ collisions  at $\sqrt{s_{NN}} = 200$ GeV from~\cite{3},~\cite{5} \label{Fig.1}}
\end{figure*}

\begin{figure*}
\caption{The relative the energy losses $\Delta p/p$ $b$ and $c$ quarks at various momenta. The upper curves
correspond to $c$ quark for $cg$ collisions(the curve 1) and for $cg + cq$ (the curve 2). The lower curves correspond
to analogous collisions for $b$ quarks (the curve 3 and  4)  \label {Fig.2}}
\end{figure*}

\begin{figure*}
\caption{The comparison of nuclear modification factor $R_{AA}^{b + c}$ in the present model at middle meaning of $p_{\perp}$
(the solid curve) with experimental data of STAR collaboration~\cite{2} for nonphotonic electrons.
 \label{Fig.3}}
\end{figure*}

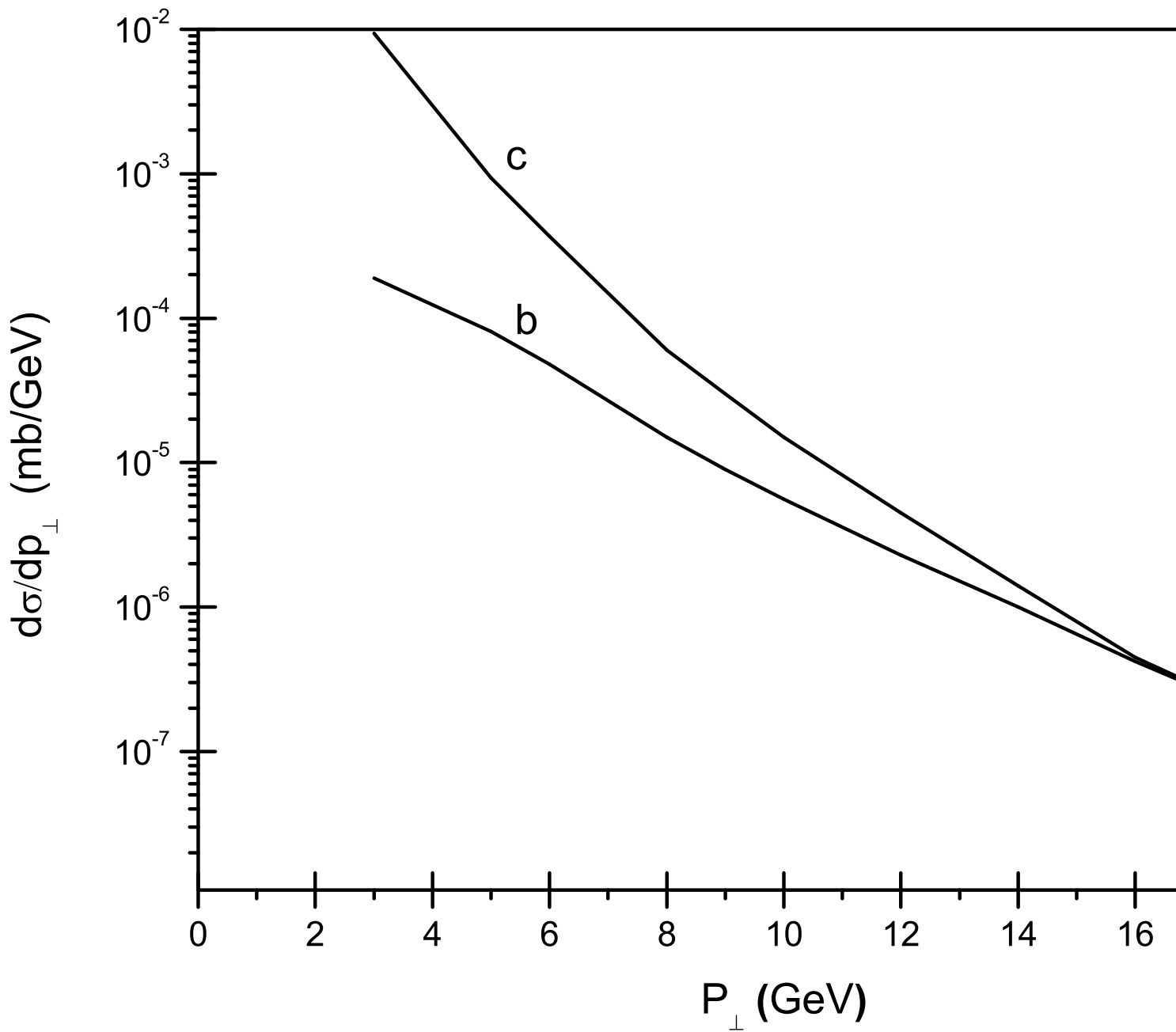
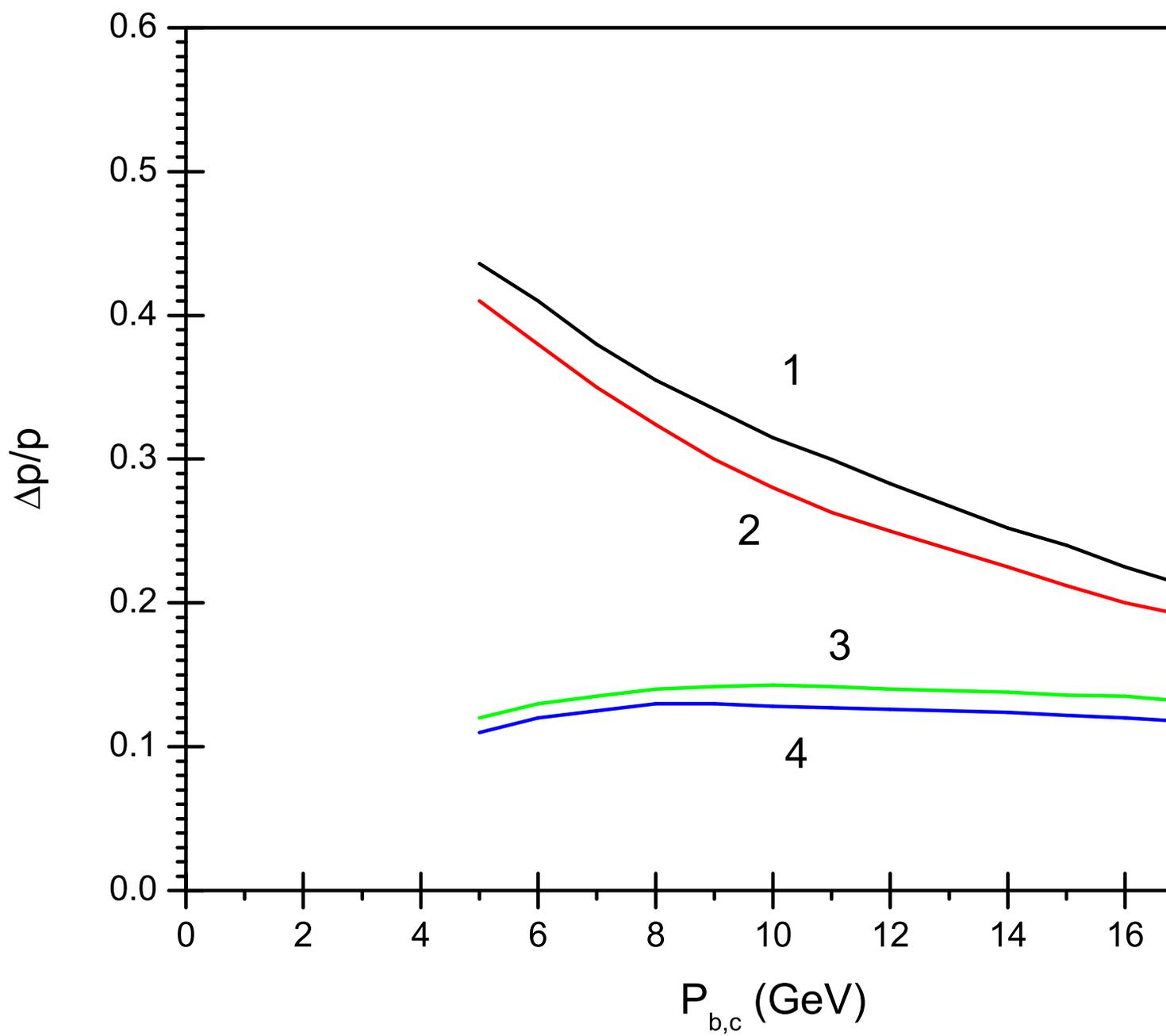
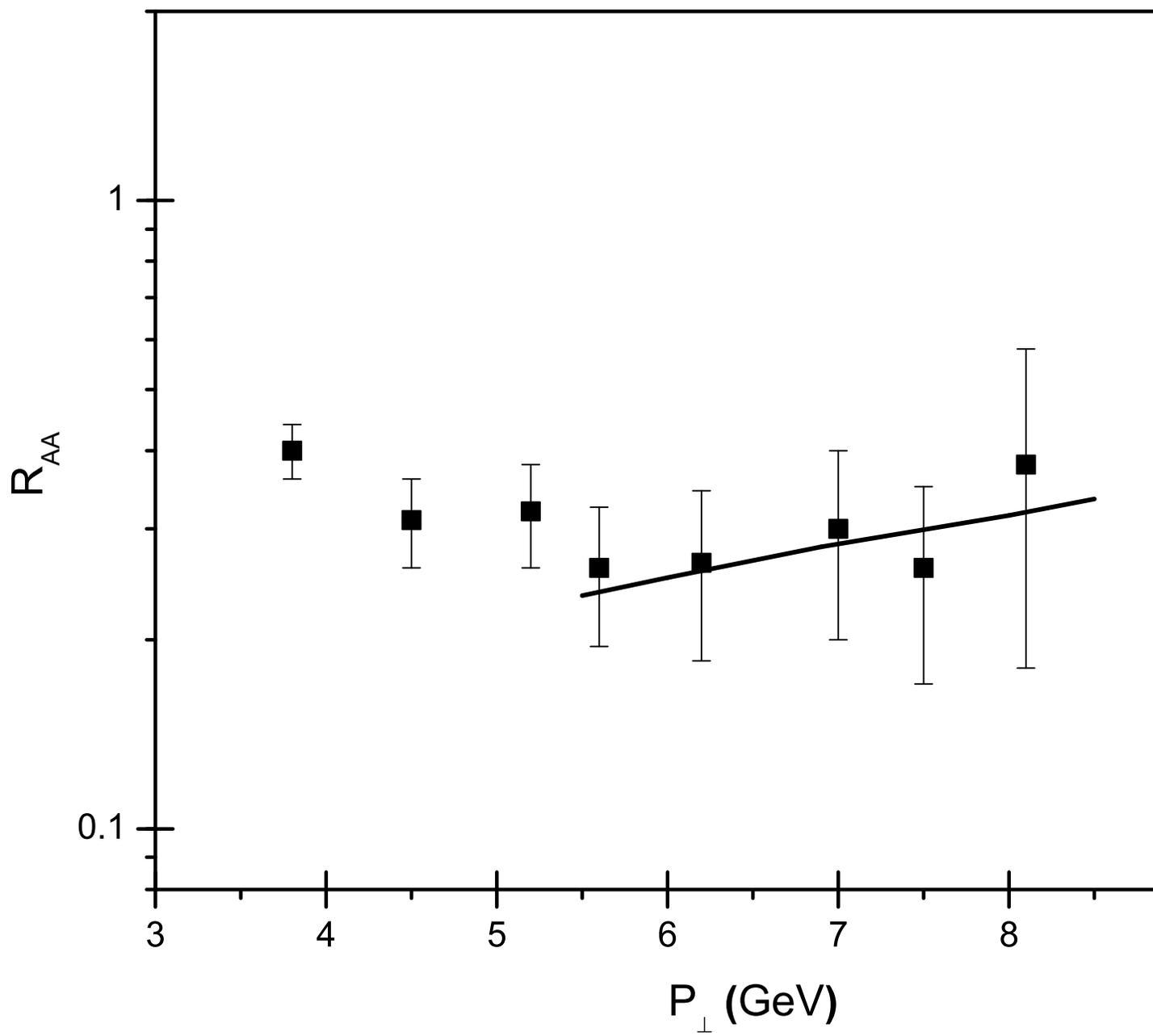
\end{document}